\documentclass{PoS}
\usepackage{amsmath}
\usepackage{verbatim}

\newcommand{\de}{\ensuremath{\delta} }
\newcommand{\De}{\ensuremath{\Delta} }
\newcommand{\ga}{\ensuremath{\gamma} }
\newcommand{\Psibar}{\ensuremath{\overline\Psi} }
\newcommand{\Qhat}{\ensuremath{\widehat Q} }
\newcommand{\muhat}{\ensuremath{\widehat\mu} }

\newcommand{\half}{\frac{1}{2}}
\newcommand{\sixt}{\ensuremath{16^3\!\times\!32} }
\newcommand{\thir}{\ensuremath{32^3\!\times\!64} }
\newcommand{\Tr}[1]{\ensuremath{\mbox{Tr}\left[ #1 \right]} }
\newcommand{\vev}[1]{\ensuremath{\left\langle #1 \right\rangle} }
\newcommand{\lsim}{\ensuremath{\lesssim} }
\newcommand{\eqn}[1]{Eqn.~\ref{#1}}
\newcommand{\fig}[1]{Fig.~\ref{#1}}
\newcommand{\refcite}[1]{Ref.~\cite{#1}}
\newcommand{\secref}[1]{Section~\ref{#1}}

\title{$S$ parameter and parity doubling \\ below the conformal window}
\ShortTitle{$S$ parameter and parity doubling below the conformal window}

\author{\speaker{David Schaich} for the Lattice Strong Dynamics (LSD) Collaboration\thanks{\texttt{http://www.yale.edu/LSD}} \\
  Physics Department \& Center for Computational Science, Boston University, Boston, MA 02215 \\
  Department of Physics, University of Colorado, Boulder, CO 80309\thanks{Present address} \\
  Email: \email{schaich@pizero.colorado.edu}
}

\abstract{Recently the Lattice Strong Dynamics Collaboration reported a reduction of the electroweak $S$ parameter for SU(3) gauge theory with $N_f = 6$ fermions in the fundamental representation, compared to scaled-up QCD.  Here I provide additional details of our calculation.  I discuss our use of conserved lattice currents; the relation to vector--axial parity doubling; finite-volume effects; and the sensitivity of our results to the number of fermion doublets with chiral electroweak couplings.  Results presented here include additional data, and do not affect our previously-published conclusions.} 

\FullConference{XXIX International Symposium on Lattice Field Theory \\ 10--16 July 2011 \\ Squaw Valley, Lake Tahoe, California}

\begin{document}
\section{Introduction} 

The application of lattice gauge theory to strongly-interacting physics beyond QCD is at present a very active field~\cite{DelDebbio:2011rc}.  While much of the current interest is motivated by the possibility that new strong dynamics may play a role in electroweak symmetry breaking~\cite{Hill:2002ap, Rychkov:2011br}, improving our general understanding of strong dynamics is an important theoretical goal in its own right.

The standard picture of strongly-interacting SU($N$) gauge theories is that as we increase the number $N_f$ of fermions in a given representation, an infrared fixed point will develop at some critical $N_f^{(c)}$.  For $N_f \geq N_f^{(c)}$ (up to the loss of asymptotic freedom) the system is IR-conformal.  Approximately-conformal systems with $N_f \lsim N_f^{(c)}$ may possess the dynamical scale separation that characterizes ``walking'' theories, as well as parity doubling between vector ($V$) and axial-vector ($A$) spectra that can reduce the electroweak $S$ parameter to phenomenologically viable values~\cite{Peskin:1991sw}.

The Lattice Strong Dynamics Collaboration approaches these questions by using QCD as a baseline.  We consider SU(3) gauge theory and steadily increase the number of fundamental fermions, comparing our results against the familiar case $N_f = 2$.  We use computationally expensive domain wall fermions for better control over lattice artifacts.  Our first studies focused on the $N_f = 6$ model, which while not truly walking exhibits some of the associated phenomena: by matching IR scales between $N_f = 2$ and $N_f = 6$ calculations, we observed an enhancement in the $N_f = 6$ chiral condensate~\cite{Appelquist:2009ka} and a reduction of the $S$ parameter relative to scaled-up QCD~\cite{Appelquist:2010xv}.  Here I provide additional details of our $S$ parameter calculation that were not discussed in \refcite{Appelquist:2010xv}.  Results presented here also include additional data, and do not affect the conclusions of \refcite{Appelquist:2010xv}.

We can identify three main ingredients in our expression for the $S$ parameter,
\begin{equation}
  \label{eq:S}
  S = 4\pi N_D \lim_{Q^2 \to 0}\frac{d}{dQ^2}\Pi_{V - A}(Q^2) - \De S_{SM}.
\end{equation}
The term $\De S_{SM}$ accounts for the three Nambu--Goldstone bosons (NGBs) eaten by the $W^{\pm}$ and $Z$, and is discussed in detail by \refcite{Appelquist:2010xv}.  In \secref{sec:currents} I review our calculation of the transverse $V$--$A$ polarization function $\Pi_{V - A}(Q^2)$, and relate it to the vector and axial spectra in \secref{sec:slopes}.  Finally, $N_D$ is the number of doublets with chiral electroweak couplings; in \secref{sec:S} I show how it affects our results for the $S$ parameter.

\section{\label{sec:currents}Currents and correlators} 

On the lattice, the transverse $V$--$A$ polarization function $\Pi_{V - A}(Q^2)$ is determined from
\begin{equation}
  \label{eq:polFunc}
  \begin{split}
    \Pi_{V - A}^{\mu\nu}(Q) & = \left(\de^{\mu\nu} - \frac{\Qhat^{\mu}\Qhat^{\nu}}{\Qhat^2}\right)\Pi_{V - A}(Q^2) - \frac{\Qhat^{\mu}\Qhat^{\nu}}{\Qhat^2}\Pi_{V - A}^L(Q^2) \\
                            & = Z\sum_x e^{iQ\cdot (x + \muhat / 2)}\Tr{\vev{\mathcal V^{\mu a}(x)V^{\nu b}(0)} - \vev{\mathcal A^{\mu a}(x)A^{\nu b}(0)}}.
  \end{split}
\end{equation}
Here $\Qhat = 2\sin(\pi n / L)$ are lattice momenta, while $Q = 2\pi n / L$; these are spacelike $Q^2 = -q^2 > 0$.  The current correlators mix two types of domain wall currents.  $V^{\mu a}$ and $A^{\mu a}$ are non-conserved ``local'' currents defined on the domain walls; in terms of five-dimensional fermion fields $\Psi(x, s)$,
\begin{equation}
  \begin{split}
    V^{\mu a}(x) & = \half\left\{\Psibar(x, L_s - 1)\ga^{\mu}(1 + \ga^5)\tau^a \Psi(x, L_s - 1) + \Psibar(x, 0)\ga^{\mu}(1 - \ga^5)\tau^a \Psi(x, 0)\right\} \\
    A^{\mu a}(x) & = \half\left\{\Psibar(x, L_s - 1)\ga^{\mu}(1 + \ga^5)\tau^a \Psi(x, L_s - 1) - \Psibar(x, 0)\ga^{\mu}(1 - \ga^5)\tau^a \Psi(x, 0)\right\}.
  \end{split}
\end{equation}
The conserved currents $\mathcal V^{\mu a}$ and $\mathcal A^{\mu a}$ are point-split, and summed over the fifth dimension:
\begin{align}
  \mathcal V^{\mu a}(x) & = \sum_{s = 0}^{L_s - 1}j^{\mu a}(x, s) &
  \mathcal A^{\mu a}(x) & = \sum_{s = 0}^{L_s - 1}\mbox{sign}\left(s - \frac{L_s - 1}{2}\right)j^{\mu a}(x, s),
\end{align}
\begin{equation}
  j^{\mu a}(x, s) = \half\left\{\Psibar(x + \muhat, s)(1 + \ga^{\mu})U_{x, \mu}^{\dag}\tau^a \Psi(x, s) - \Psibar(x, s)(1 - \ga^{\mu})U_{x, \mu}\tau^a \Psi(x + \muhat, s)\right\}.
\end{equation}
The Fourier transform in \eqn{eq:polFunc} involves $(x + \muhat / 2)$ because the conserved currents are point-split on the link $(x, x + \muhat)$.  The flavor matrices $\tau^a$ are normalized to $\Tr{\tau^a\tau^b} = \delta^{ab} / 2$.

\begin{figure}[ht]
  \centering
  \includegraphics[width=0.45\linewidth]{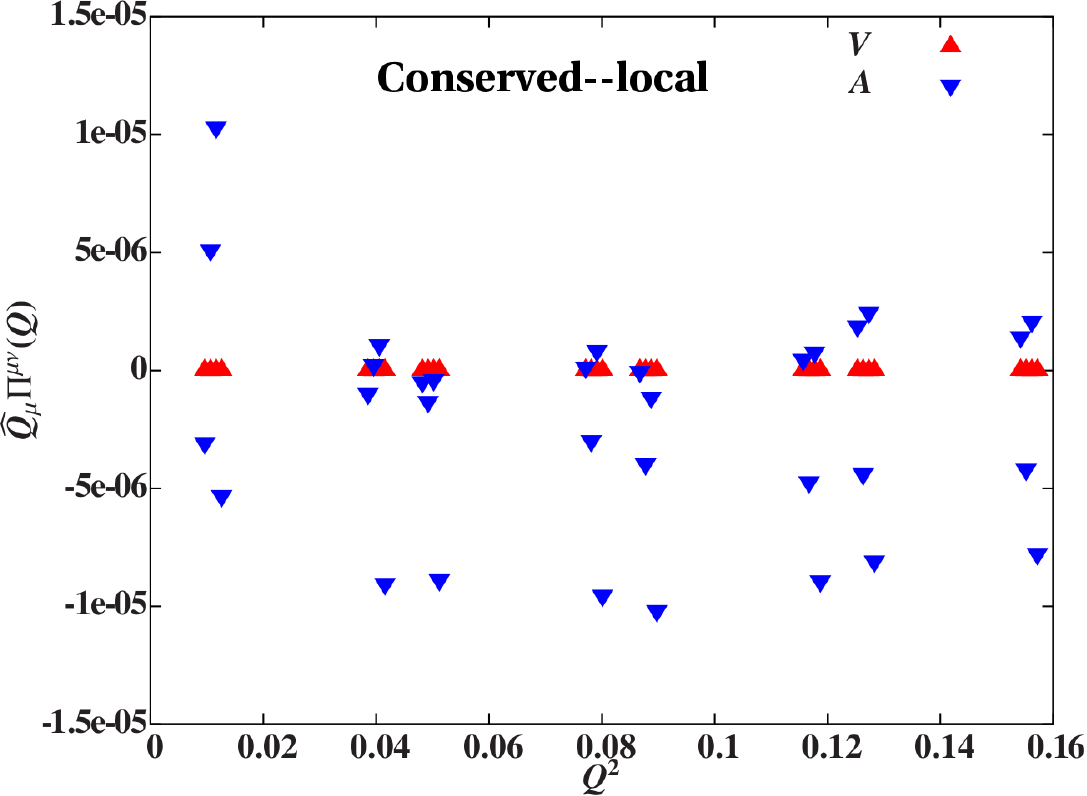}\hfill\includegraphics[width=0.45\linewidth]{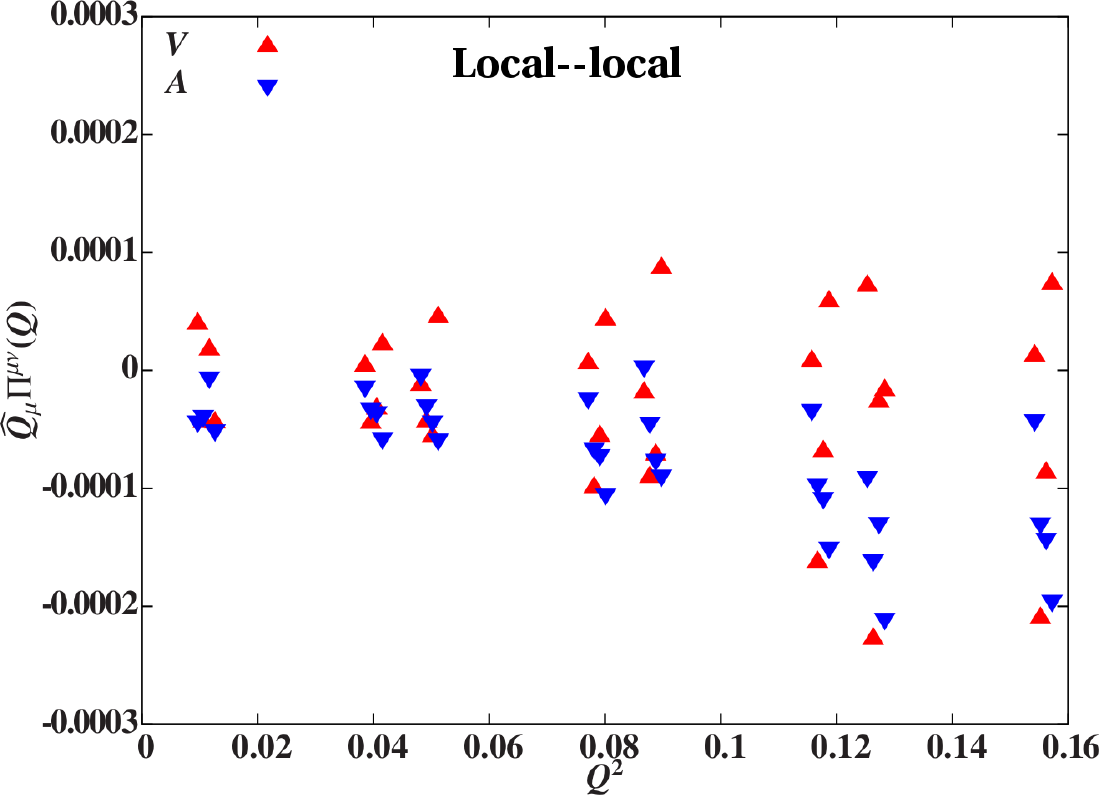}
  \caption{On every configuration, $\Qhat_{\mu} \Pi_{VV}^{\mu\nu} = 0$ when one conserved current is used in each correlator (left), but not when only non-conserved local currents are used (right).  The horizontal offsets around each $Q^2$ value distinguish different $\nu$.}
  \label{fig:Ward}
\end{figure}

Although the conserved and local currents must agree in the continuum limit, at finite lattice spacing only the former satisfy a Ward identity ($\Qhat_{\mu} \Pi_{VV}^{\mu\nu} = 0$, \fig{fig:Ward}).  Because the correlators involve both currents, \eqn{eq:polFunc} includes the renormalization factor $Z$, which we compute non-perturbatively, $Z = 0.85$ (0.73) for $N_f = 2$ (6).  Our chiral lattice fermions ensure that $Z = Z_A = Z_V$.

\begin{figure}[ht]
  \centering
  \includegraphics[width=0.45\linewidth]{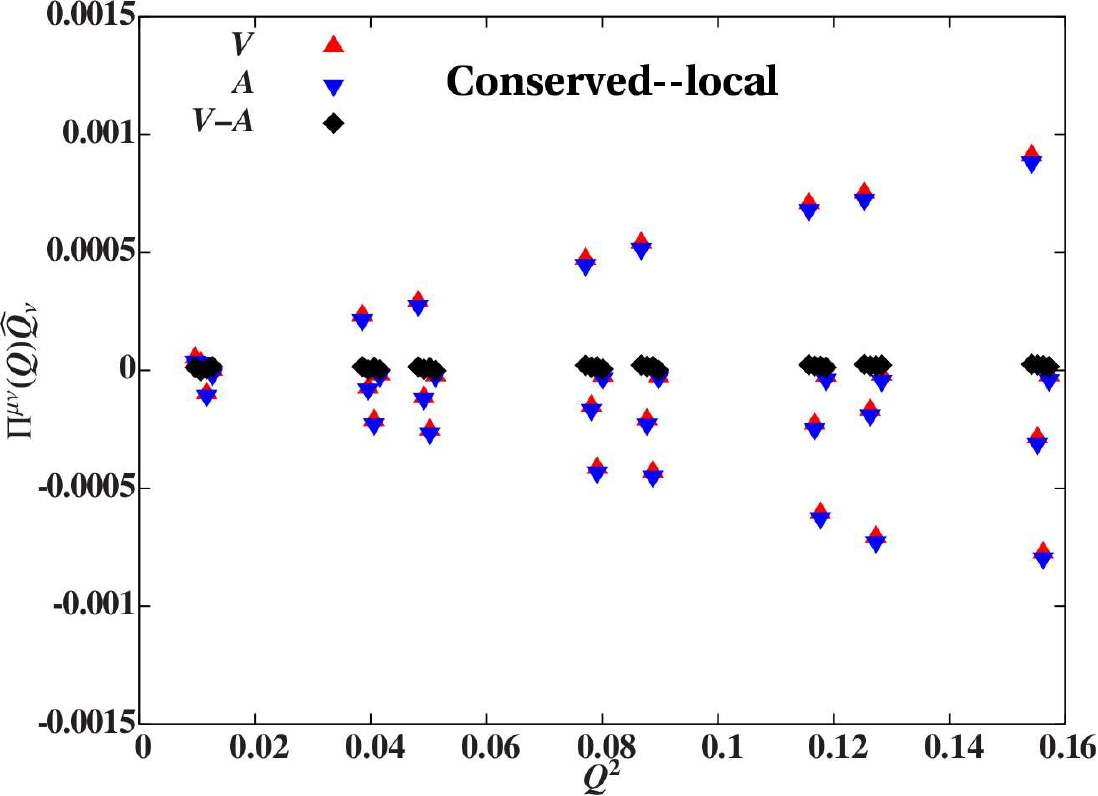}\hfill\includegraphics[width=0.45\linewidth]{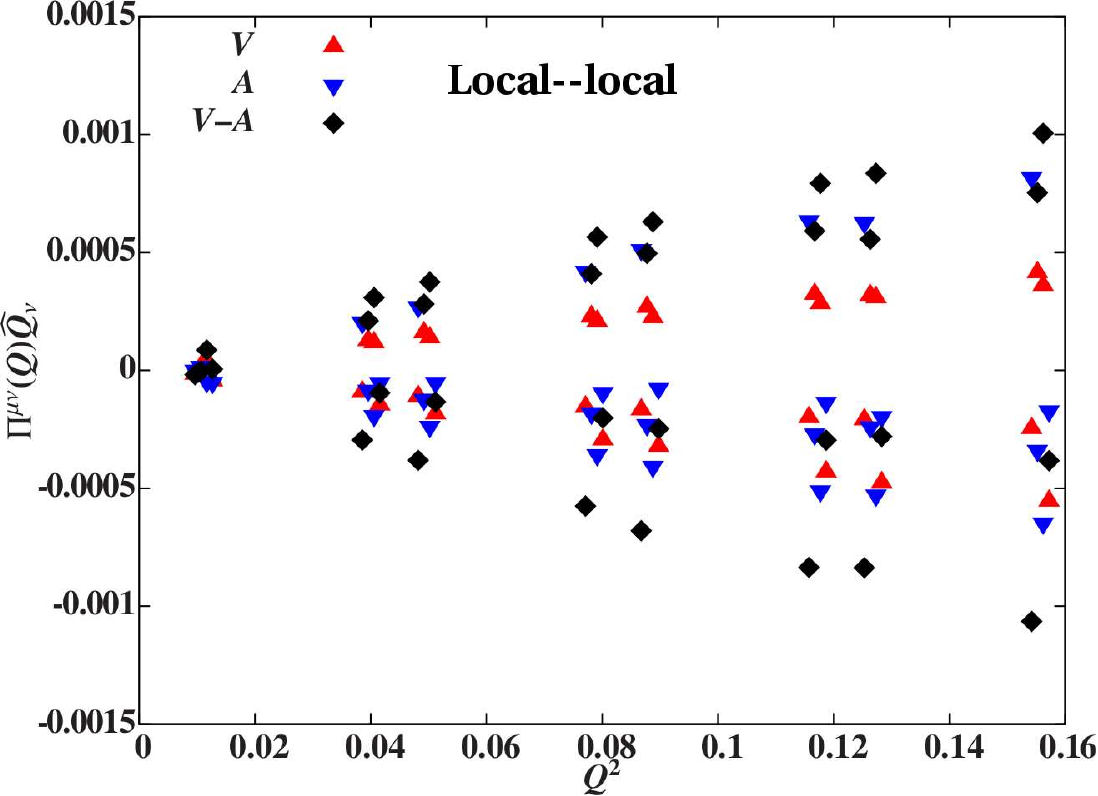}
  \caption{On every configuration, lattice artifacts $\Pi^{\mu\nu} \Qhat_{\nu} \ne 0$ cancel in the $V$--$A$ difference when one conserved current is used in each correlator (left), but not when only non-conserved local currents are used (right).  The horizontal offsets around each $Q^2$ value distinguish different $\mu$.}
  \label{fig:WardVio}
\end{figure}

In principle, it would be best to work entirely with the conserved currents $\mathcal V^{\mu a}$ and $\mathcal A^{\mu a}$ instead of using the mixed correlators in \eqn{eq:polFunc}.  In practice, evaluating conserved--conserved correlators such as $\vev{\mathcal V^{\mu a}(x) \mathcal V^{\nu b}(0)}$ requires $\mathcal O(L_s)$ inversions, increasing the computational cost of the calculation by roughly an order of magnitude.  As emphasized in \refcite{Boyle:2009xi}, lattice artifacts cancel in the $V$--$A$ difference of the mixed correlators, allowing us to use these less expensive quantities.  This is illustrated in the left panel of \fig{fig:WardVio}: even though $\Pi^{\mu\nu}\Qhat_{\nu} \ne 0$ since $V^{\nu a}$ and $A^{\nu a}$ are not conserved, $\left[\Pi_{VV}^{\mu\nu}(Q^2) - \Pi_{AA}^{\mu\nu}(Q^2)\right]\Qhat_{\nu} \approx 0$.  In the right panel, we see that this does not hold if we use only local currents in the correlators.

\section{\label{sec:slopes}Parity doubling and finite volume effects} 

Because chiral perturbation theory cannot reliably be applied to our $N_f = 6$ calculations~\cite{Neil:2010sc}, we extract the slope $\Pi_{V - A}'(0)$ by fitting our data to a simple four-parameter rational function,
\begin{equation}
  \label{eq:pade}
  \Pi_{V - A}(Q^2) = \frac{a_0 + a_1 Q^2}{1 + b_1 Q^2 + b_2 Q^4}.
\end{equation}
This ``Pad\'e(1,2)'' functional form has the correct asymptotic behavior $\Pi_{V - A}(Q^2) \sim Q^{-2}$ at large $Q^2$, and also resembles the single-pole dominance approximation to the $V$--$A$ dispersion relation
\begin{equation}
  \label{eq:dispersion}
  \Pi_{V - A}(Q^2) = -F_P^2 + \frac{Q^2}{12\pi} \int_0^{\infty} \frac{ds}{\pi} \left[\frac{R_V(s) - R_A(s)}{s + Q^2}\right].
\end{equation}
($F_P$ is the pseudoscalar decay constant.)  That is, with the single-pole dominance approximation $R(s) = 12 \pi^2 F^2 \de(s - M^2)$, this dispersion relation becomes
\begin{equation}
  \label{eq:pole}
  \Pi_{V - A}^{(pole)}(Q^2) = -F_P^2 + \frac{Q^2F_V^2}{M_V^2 + Q^2} - \frac{Q^2F_A^2}{M_A^2 + Q^2},
\end{equation}
which reproduces the form of \eqn{eq:pade} when we apply the corresponding approximation to the first Weinberg sum rule, $F_P^2 = F_V^2 - F_A^2$.  Because the lattice data contain information about the entire spectrum, the fit parameters in \eqn{eq:pade} do not directly correspond to the combinations of meson masses and decay constants predicted by the pole-dominance \eqn{eq:pole}.

\begin{figure}[ht]
  \centering
  \includegraphics[width=0.45\linewidth]{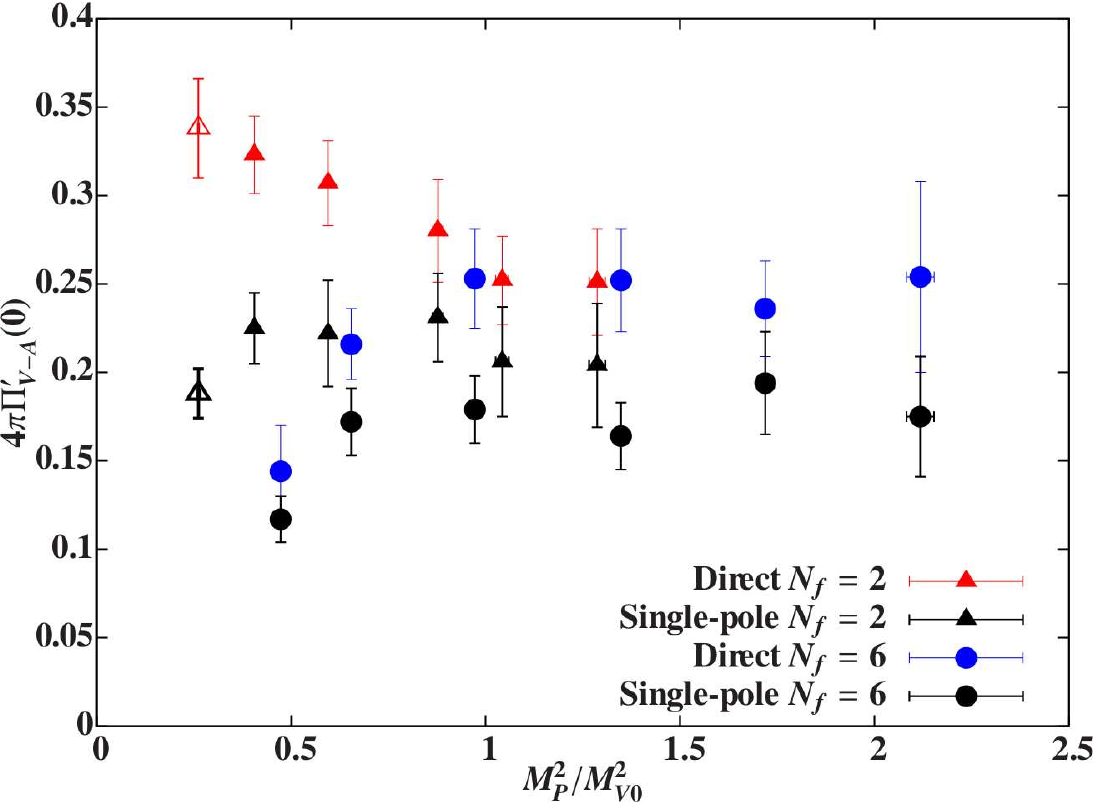}\hfill\includegraphics[width=0.45\linewidth]{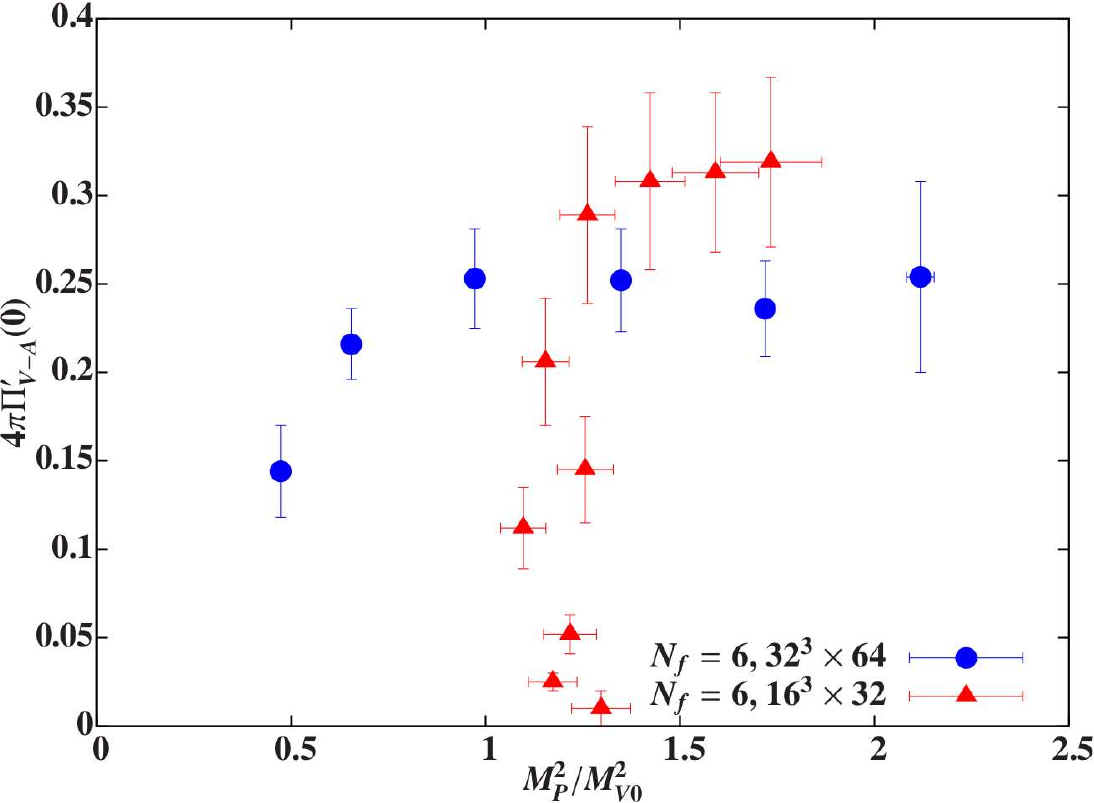}
  \caption{The slope of $\Pi_{V - A}(Q^2)$ at $Q^2 = 0$, plotted versus $M_P^2 / M_{V0}^2$.  Left: $N_f = 2$ and 6 results on \thir volumes from direct fits to Eqn.~3.1 (colored), compared to pole-dominance predictions (black).  Right: $N_f = 6$ results on \sixt and \thir volumes.}
  \label{fig:slopes}
\end{figure}

Uncorrelated fits of our data to \eqn{eq:pade} produce stable results with $\chi^2 / dof \ll 1$ as we vary the $Q^2$ fit range.  Our results for $\Pi_{V - A}'(0)$ are shown as colored points in the left panel of \fig{fig:slopes}.  The black points in that plot are pole-dominance predictions based on \eqn{eq:pole}.  Both the direct fit results and the pole-dominance predictions show a reduction for $N_f = 6$ compared to $N_f = 2$ at light pseudoscalar masses $M_P \lsim M_{V0}$, where $M_{V0}$ is the vector meson mass in the chiral limit.  The pole-dominance predictions are systematically lower than the direct results, consistent with the expectation that states neglected by the single-pole dominance approximation would provide additional positive contributions.

The lightest $N_f = 2$ points in \fig{fig:slopes} are empty because they correspond to a fermion mass $m$ so small that finite-volume effects may be significant.  Finite-volume effects are a concern for the $S$ parameter calculation because they can produce spurious parity doubling that artificially reduces $\Pi_{V - A}'(0)$.  This is illustrated in the right panel of \fig{fig:slopes} for $N_f = 6$ calculations on \sixt volumes: $\Pi_{V - A}'(0) \to 0$ as $m \to 0$, which would na\"ively suggest a negative $S$ parameter from \eqn{eq:S}.  The associated distortion of the spectrum provides clear evidence that this is merely a finite-volume effect: as $m$ decreases, the \sixt pseudoscalar mass $M_P$ freezes around $M_P^2 \approx 1.2M_{V0}^2$, which is not the case for the \thir results also shown in the plot.

Returning to the lightest $N_f = 2$ points, the pole-dominance prediction for $\Pi_{V - A}'(0)$ decreases due to spurious parity doubling from finite-volume effects.  However, we do not see a similar reduction in the direct fit result.  Instead, this point clearly continues the trend established at heavier masses, and the corresponding $N_f = 2$ results for $S$ (\fig{fig:S}, below) reproduce the prediction obtained by scaling up QCD phenomenology, $\lim_{M_P^2 \to 0} S = 0.32(3)$~\cite{Peskin:1991sw}.  This suggests that the Pad\'e fits may be less sensitive than spectral quantities to these finite-volume effects, increasing our confidence that the reduction observed for $N_f = 6$ is physical.

\section{\label{sec:S}$S$ parameter results} 

Realistic models of dynamical electroweak symmetry breaking must produce exactly three massless NGBs to be eaten by the $W^{\pm}$ and $Z$.  Any additional pseudo-Nambu--Goldstone bosons (PNGBs) must acquire masses from standard-model and other (e.g., extended-technicolor) interactions in order to satisfy experimental constraints.  On the lattice, however, we perform calculations with $N_f^2 - 1$ degenerate massive PNGBs.  When we use \eqn{eq:S} to determine the $S$ parameter from the $\Pi_{V - A}'(0)$ results shown in \fig{fig:slopes}, the $\De S_{SM}$ term removes the contribution only of the three would-be NGBs.  (To be more precise, the $I_3 = 0$ NGB does not contribute, and $\De S_{SM}$ cancels the contribution of the $|I_3| = 1$ pair.)  The remaining $N_f^2 - 4$ PNGBs introduce chiral-log terms $\propto \log[M_{V0}^2  / M_P^2]$ that would diverge in the chiral limit $M_P^2 \to 0$.

\begin{figure}[ht]
  \centering
  \includegraphics[width=0.45\linewidth]{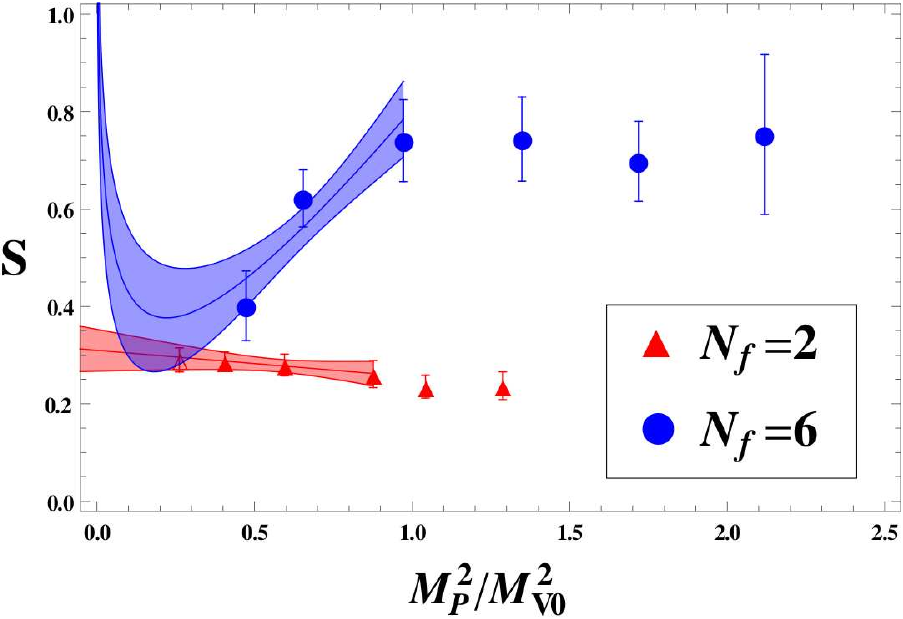}\hfill\includegraphics[width=0.45\linewidth]{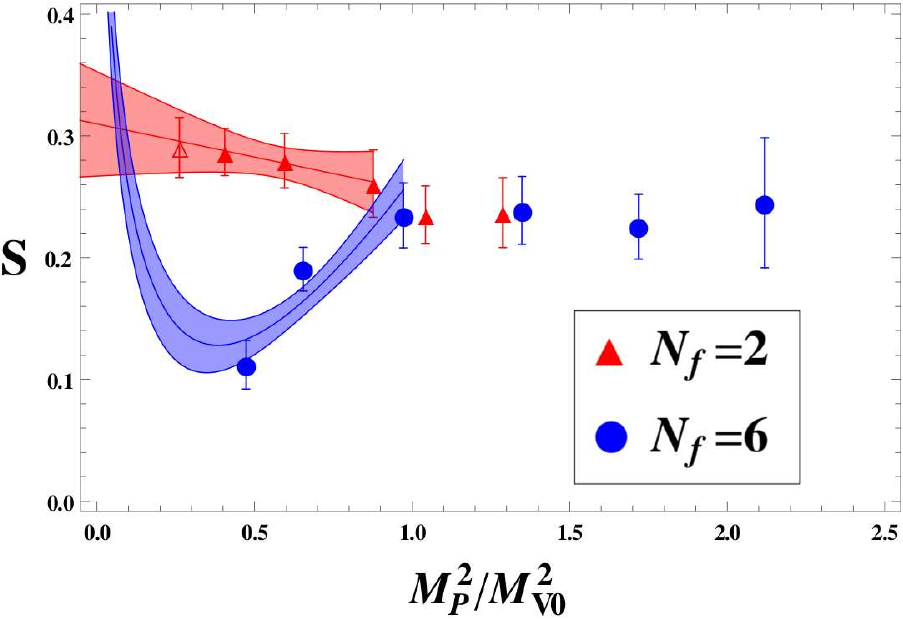}
  \caption{$S$ parameter for $N_f = 2$ and 6, for the maximum $N_D = 3$ (left) and minimum $N_D = 1$ (right).  The bands correspond to fits explained in the text.}
  \label{fig:S}
\end{figure}

\fig{fig:S} presents our $S$ parameter results for $N_f = 2$ and 6, considering two possible values of $N_D$ for $N_f = 6$.  The plot on the left presents the case in which every fermion possesses chiral electroweak couplings, $N_D = N_f / 2 = 3$.  The minimal case in which only a single doublet has chiral couplings ($N_D = 1$) is shown on the right.  In both cases the $N_f = 6$ results show a reduction compared to rescaling $N_f = 2$, before diverging in the chiral limit.  With $N_D = 1$ the $S$ parameter can be significantly closer to the experimental value $S \approx -0.15(10)$ for $M_H^{(ref)} \sim 1$ TeV~\cite{Nakamura:2010zzi}.

To guide the eye, we include in \fig{fig:S} simple linear fits accounting for the $N_D$-dependent chiral-log divergence that remains for $N_f > 2$.  We fit the lightest three solid points to the form
\begin{equation}
  S = A + Bx + \frac{\sharp - 1}{12\pi}\log\left(1 / x\right)
\end{equation}
where $x \equiv M_P^2 / M_{V0}^2$ and $\sharp$ counts the pairs of PNGBs with $I_3 \ne 0$,
\begin{align}
  \sharp & = \left(\frac{N_f}{2}\right)^2 \quad \mbox{for } N_D = N_f / 2 &
  \sharp & = 2N_f - 3 \quad \mbox{for } N_D = 1.
\end{align}
The blue $N_f = 6$ curves allow us to estimate the fermion mass $m$ at which we could directly observe chiral log effects.  The necessary $m$ is too small for us to explore on our present \thir volumes.

\begin{figure}[ht]
  \centering
  \includegraphics[width=0.45\linewidth]{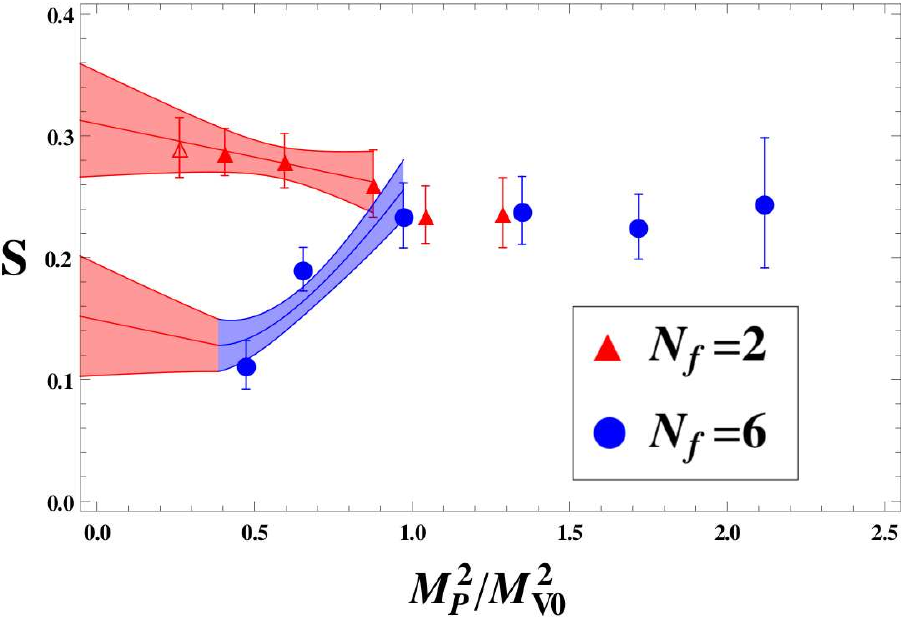}
  \caption{$S$ parameter for $N_f = 2$ and 6 with $N_D = 1$, imagining that we freeze the masses of all $N_f^2 - 4$ PNGBs at $M_P^2 = 0.38 M_{V0}^2$, as described in the text.}
  \label{fig:shenanigans}
\end{figure}

Again, in a realistic phenomenological context, we must have only three massless NGBs, with $N_f^2 - 4$ massive PNGBs.  To estimate a definite value for the $N_f = 6$ $S$ parameter in this situation, we can imagine freezing the masses of all $N_f^2 - 4$ PNGBs at some finite value (such as $M_P^2 = 0.38 M_{V0}^2$ at the minimum of the $N_D = 1$ blue curve in \fig{fig:S}), and then taking only the three NGBs to the chiral limit $M_P^2 \to 0$.  A qualitative picture of this scenario is sketched in \fig{fig:shenanigans}.

\section*{Acknowledgments} 

I thank the members of the LSD Collaboration for many useful discussions, and review of this contribution: T.~Appelquist, R.~Babich, R.~Brower, M.~Buchoff, M.~Cheng, M.~Clark, S.~Cohen, G.~Fleming, J.~Kiskis, M.~Lin, H.~Na, E.~Neil, J.~Osborn, C.~Rebbi, S.~Syritsyn, P.~Vranas, G.~Voronov, J.~Wasem and O.~Witzel.  This work was supported by the U.S.~Department of Energy (DOE) through grants DE-FG02-91ER40676 and DE-FG02-04ER41290; the Lawrence Livermore National Laboratory Institutional Computing Grand Challenge program; the DOE Scientific Discovery through Advanced Computing program through the USQCD Collaboration;\footnote{\texttt{http://www.usqcd.org}} the U.S.~National Science Foundation through TeraGrid resources provided by the National Institute for Computational Sciences under grant number TG-MCA08X008;\footnote{\texttt{http://www.xsede.org}} and Boston University's Scientific Computing Facilities.

\bibliographystyle{utphys}
\bibliography{Lattice2011_proc}
\end{document}